# Nuclear Resonance Vibrational Spectroscopy of Iron-Sulfur Proteins

A.X. Trautwein, P. Wegner, H. Winkler, H. Paulsen
*Institut für Physik, Universität zu Lübeck, D-23538 Lübeck, Germany*

V. Schünemann
*Fachbereich Physik, Technische Universität Kaiserslautern, D-67663 Kaiserslautern, Germany*
`(schuene@physik.uni-kl.de)`

C. Schmidt
*Institut für Biochemie, Universität zu Lübeck, D-23538 Lübeck, Germany*

A.I. Chumakov, R. Rüffer
*ESRF, F-38043 Grenoble Cedex 9, France*

**Abstract.** *Nuclear inelastic scattering in conjunction with density functional theory (DFT) calculations has been applied for the identification of vibrational modes of the high-spin ferric and the high-spin ferrous iron-sulfur center of a rubredoxin-type protein from the thermophylic bacterium Pyrococcus abysii.*

## 1. Introduction

Iron-sulfur proteins function as electron-transfer proteins in all living cells. They are involved in photosynthesis, cell respiration as well as in nitrogen fixation. Most iron-sulfur proteins have either single-, two-, three- or four-iron centers. Mössbauer spectroscopy has contributed a lot to the understanding of the electronic ground states of these clusters [1]. The vibrational properties of the iron centers, however, which are related to their biological function, are much less studied. This is partly due to the fact that the vibrational states of the iron centers are masked by the vibrational states of the protein backbone and thus conventional spectroscopy like Resonance Raman or infrared spectroscopy does not provide a clear picture of the vibrational properties of the iron centers. Nuclear inelastic scattering of synchroton radiation (NIS) measurements allow to identify individual iron modes and therefore are sensitive to iron-ligand bonds which are in turn related to the formal oxidation state of the iron.

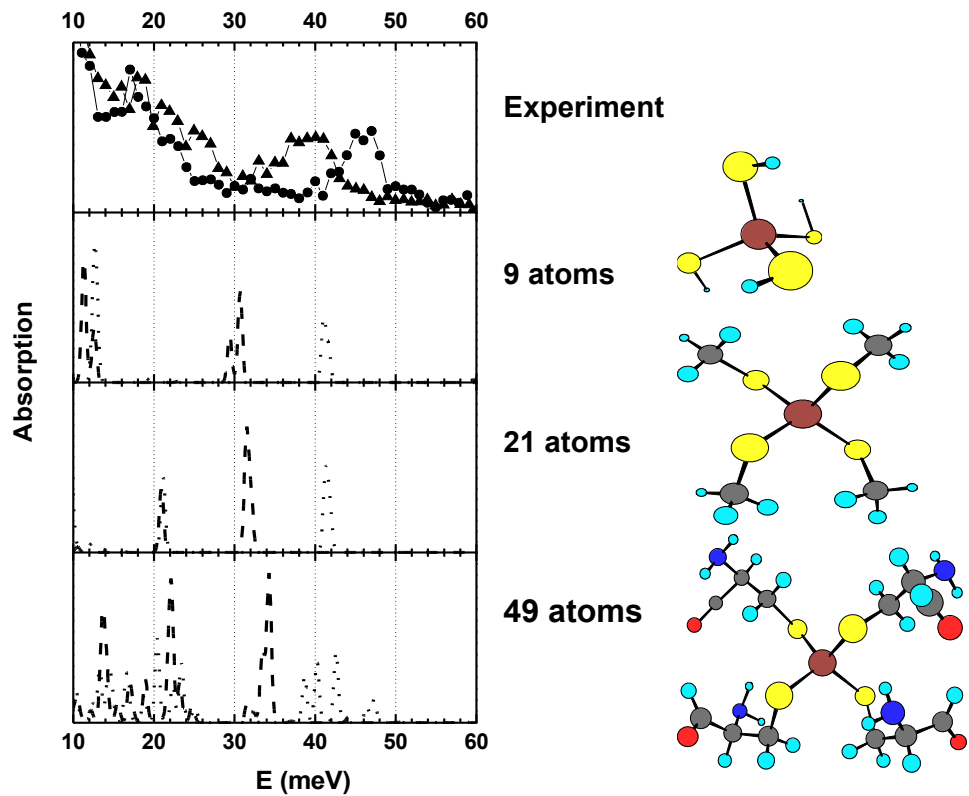

**Figure 1**. NIS-spectra of oxidized (●) and reduced rubredoxin mutant Rm 2-4 (▲) from *Pyrococcus abyssi* obtained at 25 K. Theoretically calculated NIS spectra based on DFT calculations (B3LYP/CEP-31G) of 9, 21 and 49 atoms are shown below. The dashed lines represent calculated NIS spectra for the oxidized $Fe^{III}$-$S_4$ center and the dotted line for the reduced $Fe^{II}$-$S_4$ center.

## 2. Method

NIS measurements have been performed on the rubredoxin type iron-sulfur protein mutant Rm 2-4 from *Pyrococcus abysii* [2], which has a single Fe-$S_4$ center, at the beamline ID 18 of ESRF in Grenoble, France at T = 25 K. Protein samples have been prepared with $^{57}$Fe concentrations of *up to* 10 mM. DFT calculations were performed for three different models (Fig. 1) *in vacuo* with the B3LYP hybrid functional [3] together with the CEP-31G basis [4] as included in the Gaussian 03 program package [5].

### 3. Results and Discussion

Figure 1 shows NIS measurements on the oxidized $Fe^{III}$-$S_4$ protein which reveal a broad symmetric band around 15-25 meV (121-202 cm$^{-1}$) and a broad asymmetric band around 42-48 meV (339-387 cm$^{-1}$) consistent with the results of Bergmann et al. on rubredoxin from *Pyrococcus furiosus* [6]. These results confirm Resonance Raman studies on oxidized rubredoxin from *desulfovibrio gigas* which indicates that there are three bands (43.15, 45.01 and 46.62 meV; 348, 363 and 376 cm$^{-1}$) in the region where asymmetric $Fe^{III}$-S stretch modes are expected [7,8]. Resonance Raman data of the reduced rubredoxin with its iron in the $Fe^{II}$-$S_4$ (S=2) state could not be obtained up to now [8]. Thus NIS is the method of choice to study the dynamical properties of $Fe^{II}$-$S_4$ centers. Upon reduction the asymmetric stretch modes observed by NIS shift to lower energies (36-42 meV; 291-339 cm$^{-1}$). This can be rationalized by the fact that upon reduction the $Fe^{II}$-S bond lengths increase, which in turn is accompanied by a decrease of $Fe^{II}$-S binding energy and therefore by a decrease of the force constant of the $Fe^{II}$-S bond. This correlation between iron-oxidation state and Fe-S stretch-mode position is in line with DFT calculations for three different models (Fig. 1), which have been performed on the basis of the obtained crystal structure of the protein. The results for the model with 21 atoms are shown in Table 1. NIS visible S-Fe-S bending modes are expected in the region from 6-10 meV, Fe-S-C bending modes from 16-21 meV and Fe-S streching modes around 28-42 meV.

However, although the shift of the asymmetric stretch modes is well reflected by the DFT calculation, the line positions are less well reproduced. This is due to the fact that obviously more than 21 and even more than 49 atoms (see Fig. 1) have to be regarded in the DFT calculations.

In conclusion this NIS study has provided first information about the vibrational properties of a reduced iron-sulfur center in a protein. These studies demonstrate that it is profitable to apply nuclear inelastic scattering as a complementary method to IR and Raman spectroscopy for investigating the dynamical properties of iron sites in iron sulfur proteins.

**Acknowledgement**

This work has been supported financially by ESRF (experiment SC-1217).

Table 1. Assignment of vibrational iron-ligand modes based on a DFT calculation on the model with 21 atoms (Fig. 1).

| Character of vibrational modes | Energy (meV) | |
|---|---|---|
| | $Fe^{III}$-$S_4$ | $Fe^{II}$-$S_4$ |
| 4 torsional modes | 1-6 | 2-5 |
| 2 S-Fe-S bending modes   (NIS vis.) | 8-10 | 6-8 |
| 7 torsional modes | 10-14 | 9-13 |
| 4 Fe-S-C bending modes   (NIS vis.) | 17-21 | 16-21 |
| 4 Fe-S stretching modes   (NIS vis.) | 35-42 | 28-32 |